\documentclass [11pt,a4paper] {article}

\usepackage[T1]{fontenc}
\usepackage{lmodern}

\usepackage[cp1252]{inputenc}
\usepackage{amssymb}
\usepackage{amsmath}
\usepackage{amsfonts,amssymb}

\usepackage{graphicx}

\usepackage{bbm}
\usepackage{enumerate}

\usepackage[shortlabels]{enumitem}

\usepackage{amsthm}
\usepackage{cancel}

\usepackage{centernot}
\usepackage{mathrsfs}

\DeclareMathAlphabet{\mathpzc}{OT1}{pzc}{m}{it}

\begin{document}

\title{Resolving the trans-Planckian problem along the lines of a finite geometry}

\author{Arkady Bolotin\footnote{$Email: arkadyv@bgu.ac.il$\vspace{5pt}} \\ \emph{Ben-Gurion University of the Negev, Beersheba (Israel)}}

\maketitle

\begin{abstract}\noindent In black hole physics, inflationary cosmology, and quantum field theories, it is conjectured that the physical laws are subject to radical changes below the Planck length. Such changes are due to effects of quantum gravity believed to become significant at the Planck length. However, a complete and consistent quantum theory of gravity is still missing, and candidate models of quantum gravity have not yet overcome major formal and conceptual difficulties. Another problem is how to determine a geometry of physical space that features a minimal length scale such as the Planck length. In the present paper it is demonstrated that the said geometry can be any geometric system omitting continuity, i.e., a geometry that possesses only a finite number of points.\bigskip\bigskip

\bigskip\bigskip
\end{abstract}

\section{Introduction}  

\noindent Does nature feature a minimal length scale $\ell_{\text{min}}$? Is it true to say that $\ell_{\text{min}}$ coincides with the Planck length $\ell_{P}$? On account of the generalized uncertainty principle (GUP) \cite{Mead, Adler, Hossenfelder} and the holographic principle (HP) \cite{Hooft, Susskind, Bigatti}, one may incline to think that both those questions have a positive answer.\bigskip

\noindent Consider for example a thought experiment proposed in \cite{Salecker} that concerns the registration of photons reflected off a mirror at some distance $D$ from a non-relativistic detector. Let the variance of the position of the detector $\ell \ge 0$ be $\mathrm{Var}(\ell)$. Then, in accordance with the Heisenberg uncertainty principle, the variance of the detector’s velocity must be $\mathrm{Var}(\dot{\ell})\ge \hbar^2/4\mathrm{Var}(\ell)M^2$, where $M$ is the detectors's mass. Observe that the time needed for a photon to travel to the mirror and come back is $T=2D/c$. Let the detector have moved by $\dot{\ell}T$ during that time. On the assumption that $\ell$ and $\dot{\ell}T$ are uncorrelated random variables, the variance of their sum must be equal to the sum of their variances, or, expressed symbolically: $\mathrm{Var}(\ell + \dot{\ell}T) = \mathrm{Var}(\ell)+\mathrm{Var}(\dot{\ell})T^2$. Provided that the registration of photons reflected off the mirror will not be connected to the rest of the world if the distance from the detector to the mirror $D$ is closer than or equal to the Schwarzschild radius $2GM/c^2$, one finds:\smallskip

\begin{equation}  
   \mathrm{Var}\left(\ell + \dot{\ell}T\right)
   \ge
   \mathrm{Var}(\ell)
   +
   \frac{4\ell_{P}^4}{\mathrm{Var}(\ell)}
   \;\;\;\;   ,
\end{equation}
\smallskip

\noindent where $\ell_{P}^4=\hbar^2 G^2/c^6$. The above is a generalization of the uncertainty principle that accounts for gravitational effects in the measure of positions. The variance $\mathrm{Var}(\ell + \dot{\ell}T)$ gets the minimum when $\mathrm{Var}(\ell)=\min{\!(\mathrm{Var}(\ell))}=2\ell_{P}^2$. As $\min{\!(\mathrm{Var}(\ell))} \sim \ell_{\text{min}}^2$, this entails the lower bound on the variable $\ell$, namely, $\ell_{\text{min}}=\sqrt{2}\ell_{P} \approx \ell_{P}$.\bigskip

\noindent Regarding HP, it states that physics inside a bounded region is fully captured by physics at the boundary of the region \cite{Bousso}. As a consequence, the vacuum enclosed inside a region with a boundary of area $A$ is fully described (up to a factor of $\log{2}$) by no more than $A/\ell_{P}^2$ degrees of freedom. Stipulating $A_{\text{min}}/\ell_{P}^2=1$ and $A_{\text{min}} \sim \ell_{\text{min}}^2$, this implies $\ell_{\text{min}} \sim \ell_{P}$.\bigskip

\noindent On the other hand, neither GUP nor HP has implications for a geometry of physical space. In more detail, since GUP is expressed in terms of variances, there is a nonzero chance of $\ell$ being within the interval $\ell < \ell_{P}$. Additionally, HP relates to discretization, i.e., the process of transferring continuous geometrical magnitudes such as areas into sets of primitive objects, e.g., Planck areas $\ell_{P}^2$. As such, it prompts to discretization error that satisfies the equation:\smallskip

\begin{equation}  
   A
   <
   \Big\lfloor\, \frac{A}{\ell_{P}^2} \,\Big\rfloor
   \ell_{P}^2
   +
   \ell_{P}^2
   \;\;\;\;   .
\end{equation}
\smallskip

\noindent Denoting the floor function $\lfloor {A}/{\ell_{P}^2} \rfloor$ by $m \in \mathbb{N}$ and supposing $A=m\ell_{P}^2+\ell^2$ indicates that there are distances $\ell < \ell_{P}$. Hence, despite GUP and HP, a geometry of physical space continues to be without limits or bounds. A case in point is a gravitational singularity where all spatial dimensions become of size zero.\bigskip

\noindent The appearance of distances beyond the Planck length presents a problem (known as the trans-Planckian problem \cite{Jacobson, Brout}) because one expects the physical laws to undergo fundamental changes beyond $\ell_{P}$. To address the problem, one may propose that gravitational singularities do not exist. The idea can be stated in the form that due to effects of quantum gravity, there is a significant deviation from $1/\ell^2$  law of Newtonian gravity beyond the minimum distance $\ell_{\text{min}} = \ell_{P}$. Consequently, distances $\ell < \ell_{P}$ are unreachable by way of the gravitational force.\bigskip

\noindent Be that as it may, a complete and consistent quantum theory of gravity is still missing, and candidate models of quantum gravity still need to overcome major formal and conceptual problems \cite{Kiefer}. Giving that, alternative ways of preventing distances $\ell < \ell_{P}$ from appearing deserve to be considered and analyzed.\bigskip

\noindent One of those ways is to assume that a geometry of physical space excludes continuum, that is, the number of points that make up physical space is finite. This approach will be discussed in the present paper.\bigskip

\section{A finite plane geometry}  

\noindent A geometry can be defined as a system of axioms which identify what things are that constitute fundamental objects such as points and lines \cite{Bezdek}. In terms of this definition, a finite geometry is any of axiomatic systems which permit only a finite number of points.\bigskip

\noindent Suppose that $\mathbb{M}^2$ is a plane (i.e., a flat, two-dimensional space) that can be called a finite geometry and $\mathcal{A}$ is a region in $\mathbb{M}^2$ whose area is $A$. Let the cardinality of $\mathcal{A}$ (i.e., the number of points that constitute $\mathcal{A}$) be denoted by $P(\mathcal{A})$ and let it conform to the area $A$ expressed in units of the Planck area:\smallskip

\begin{equation}  
   \mathrm{card}(\mathcal{A})
   \equiv
   P(\mathcal{A})
   \sim
   \frac{A}{\ell_{P}^2}
   \;\;\;\;  .
\end{equation}
\smallskip

\noindent Consider $\mathcal{C}(\mathcal{A})$, the system of coordinates on $\mathcal{A}$, that is, the set of numbers that specify the position of each point in $\mathcal{A}$. Since $\mathbb{M}^2$ is finite, the region $\mathcal{A}$ is finite and so is the set $\mathcal{C}(\mathcal{A})$. Moreover, to be qualified as a system of coordinates, the set $\mathcal{C}(\mathcal{A})$ must be such that any binary operation on its elements is defined. More precisely, a mapping\smallskip

\begin{equation}  
   f\!\!:
   \mathcal{C}(\mathcal{A}) \times \mathcal{C}(\mathcal{A})
   \to
   \mathcal{C}(\mathcal{A})
   \;\;\;\;    
\end{equation}
\smallskip

\noindent is required to exist (meaning that binary operations are supposed to be closed on $\mathcal{C}(\mathcal{A})$). This implies that the set $\mathcal{C}(\mathcal{A})$ must be a finite field $\mathbb{F}_q$ whose size is a prime power $q = p^n$ with a prime number $p$ and a positive integer $n$ \cite{Mullen13}. Therefore, the cardinality of $\mathcal{C}(\mathcal{A})$ can be determined as\smallskip

\begin{equation}  
   \mathrm{card}\left( \mathcal{C}(\mathcal{A}) \right)
   =
   \mathrm{card}(\mathbb{F}_q)
   =
   p^{m P(\mathcal{A})}
   \;\;\;\;  .
\end{equation}
\smallskip

\noindent where $m \in \{1,2,\dots\}$. This system of coordinates can store\smallskip

\begin{equation}  
   \log_2 \mathrm{card}\left( \mathcal{C}(\mathcal{A}) \right)
   =
   m \, P(\mathcal{A}) \log_2 p
   \;\;\;\;   
\end{equation}
\smallskip

\noindent bits of information ($m \log_2 p$ bits per each point in $\mathcal{A}$). Picking out $p$ and $m$ to be 2 and 1 respectively, i.e., choosing the size of $\mathcal{A}$ to be\smallskip

\begin{equation} \label{SCOOR} 
   \mathrm{card}\left( \mathcal{C}(\mathcal{A}) \right)
   =
   2^{P(\mathcal{A})}
   \;\;\;\;  ,
\end{equation}
\smallskip

\noindent allows one to construe each point in $\mathcal{A}$ as a bit of information.\bigskip

\noindent To make our discussion more tangible, let us assume that $\mathcal{A}$ consists of just 4 points, i.e., $P(\mathcal{A})=4$. Then, $\mathrm{card}\left( \mathcal{C}(\mathcal{A}) \right) = 2^4$ or\smallskip

\begin{equation} \label{PLANE} 
   \mathrm{card}\left( \mathcal{C}(\mathcal{A}) \right)
   =
   \mathrm{card}(\mathbb{F}_4)^2
   \;\;\;\;  ,
\end{equation}
\smallskip

\noindent where $\mathrm{card}(\mathbb{F}_4)^2$ is the cardinality of a vector space of dimension 2 over the finite field $\mathbb{F}_4$.\bigskip

\noindent The field $\mathbb{F}_4$ consists of four elements called $O$, $I$, $\alpha$ and $\beta$ \cite{Mullen07}. Therewithal, $O$ plays the role of the additive identity element, 0, at the same time as $I$ fulfils the role of the multiplicative identity element, 1. Furthermore, $\alpha^2=\beta$, $\beta^2=\alpha$, and\smallskip

\begin{equation}  
   \forall x \in \{ O,I,\alpha,\beta\}\!\!:\;\;
   x+x = O
   \;\;\;\;   .
\end{equation}
\smallskip

\noindent The expression (\ref{PLANE}) implies that the elements of $\mathcal{C}(\mathcal{A})$ are the points of the affine plane $\mathbb{F}_{16}=\mathbb{F}_4\times\mathbb{F}_4$. These points are identified with ordered pairs $(x,y)$ of numbers modulo 4 and can be connected with lines satisfying a linear equation\smallskip

\begin{equation}  
   ax+by = c \; (\mathrm{mod}\; 4)
   \;\;\;\;   .
\end{equation}
\smallskip

\begin{figure}[ht!]
   \centering
   \includegraphics[scale=0.52]{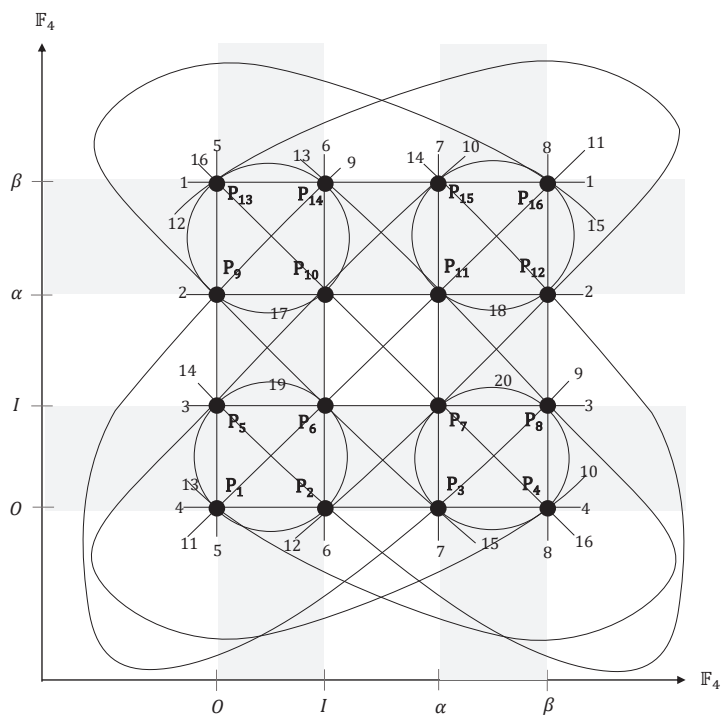}
   \caption{The affine plane $\mathbb{F}_4\times\mathbb{F}_4$.\label{fig1}}
\end{figure}

\noindent In this way, $\mathcal{C}(\mathcal{A})$ can be presented as the configuration in the plane of 16 points and 20 (straight) lines. The Fig.1 shows this configuration (the lines 9, 10, 12-15 are drawn as curves and the lines 17-20 are pictured as circles). In the language of configuration, the one shown in the Fig.1 has the notation $16_5 20_4$ (meaning that there are 16 points, 5 lines per point, 20 lines, and 4 points per line). The affine plane displayed in the Fig.1 has no ordinary lines (i.e., ones that contain exactly two of the set of points) but possesses “parallel” lines (i.e., ones that have no common points); for example, the line 11 is parallel to the line 16.\bigskip

\noindent The configuration $16_5 20_4$ is known as Sylvester-Galai configuration (SGC) \cite{Boros}. It cannot be realized by points and lines of the Euclidean plane. This suggests that a geometry based on SGC is not metric, to be specific, such a geometry cannot be equipped with the Euclidean distance satisfying all the metric axioms.\bigskip

\noindent The last can be demonstrated explicitly.\bigskip

\section{The Euclidean distance on an affine geometry}  

\noindent Given that subtraction is identical to addition, as is the case for every filed $\mathbb{F}_2^m$ with $m\in\{1,2,\dots\}$, one can introduce the distance $d(x_1,x_2)$ between distinct elements $x_1$ and $x_2$ in $\mathbb{F}_4$ by the formula\smallskip

\begin{equation}  
   \forall x_1,x_2 \in \mathbb{F}_4\!\!:\;\;
   d(x_1,x_2) = x_1 + x_2
   \;\;\;\;   .
\end{equation}
\smallskip

\noindent Such a distance is metric because it satisfies all the metric axioms, which are:\smallskip

\begin{description}[itemindent=+0.2cm]
\item[$\mathbf{(M1)}$] \text{Identity of indiscernibles:}\, $\forall x_1,x_2 \in \mathbb{F}_4 \,\Big( x_1=x_2 \to d(x_1,x_2)=O \Big),$
\item[$\mathbf{(M2)}$] \text{Positiveness:}\,                  $\forall x_1,x_2 \in \mathbb{F}_4 \,\Big( x_1 \neq x_2 \to d(x_1,x_2) \in \mathbb{F}_4\setminus O \Big),$
\item[$\mathbf{(M3)}$] \text{Symmetry:}\,                     $\forall x_1,x_2 \in \mathbb{F}_4 \,\Big( d(x_1,x_2) = d(x_2,x_1) \Big),$
\item[$\mathbf{(M4)}$] \text{Triangle inequality:}\,         $\forall x_1,x_2,x_3 \in \mathbb{F}_4 \,\Big( d(x_1,x_3) = d(x_1,x_2)+d(x_2,x_3) \Big).$
\end{description}

\noindent Certainly, those axioms can be easily verified by replacing the distance between two elements in $\mathbb{F}_4$ for their total. For example, since $x_2+x_2=O$, one finds\smallskip

\begin{equation}  
   x_1 + x_3
   =
   (x_1 + x_2)
   +
   (x_2 + x_3)
   =
   x_1 + O + x_3
   \;\;\;\;   .
\end{equation}
\smallskip

\noindent The above means to imply that the ordered pair $(\mathbb{F}_4, x_1+x_2)$ is a metric space over the field $\mathbb{F}_4$.\bigskip

\noindent Following the treatment of \cite{Burago}, consider the direct product of two metric spaces over $\mathbb{F}_4$\smallskip

\begin{equation} \label{PROD} 
   (\mathbb{F}_4, x_1+x_2)
   \times
   (\mathbb{F}_4, y_1+y_2)
   =
   \Big(
      \mathbb{F}_4 \times \mathbb{F}_4
      ,
      (x_1+x_2) \times (y_1+y_2)
   \Big)
   \;\;\;\;  ,
\end{equation}
\smallskip

\noindent where the distance between two points on the plane $\mathbb{F}_{16} =\mathbb{F}_4 \times \mathbb{F}_4$ with coordinates $(x_1,y_1)$ and $(x_2,y_2)$ is the Euclidean distance\smallskip

\begin{equation}  
   d\left((x_1,y_1),(x_2,y_2)\right)
   =
   (x_1+x_2) \times (y_1+y_2)
   =
   \sqrt{(x_1+x_2)^2 + (y_1+y_2)^2}
   \;\;\;\;  .
\end{equation}
\smallskip

\noindent In view of (\ref{PROD}), the set $\mathcal{C}(\mathcal{A})$ is the coordinate system that specifies every point in $\mathcal{A}$ by a pair of the elements $(x,y)$ in $\mathbb{F}_{16}$ which are the distances $x+O=x$ and $y+O=y$ to the point from two fixed coordinate lines.\bigskip

\noindent Therefore, $(x,y)$ are Cartesian coordinates of points in $\mathcal{A}$ and, correspondingly, $\mathcal{C}(\mathcal{A})$ is the Cartesian coordinate system for $\mathcal{A}$.\bigskip

\noindent By virtue of the identity\smallskip

\begin{equation}  
   \forall x_1,x_2 \in \mathbb{F}_4\!\!:\;\;
   (x_1 + x_2)^2
   =
   x_1^2 + x_2^2
   \;\;\;\;    
\end{equation}
\smallskip

\noindent the Euclidean distance on the plane $\mathbb{F}_{16}$ takes the form of sum\smallskip

\begin{equation}  
   d\left((x_1,y_1),(x_2,y_2)\right)
   =
   \underbrace{\sqrt{(x_1 + x_2)^2+(y_1+y_2)^2}}_{\sqrt{a^2+b^2}=\sqrt{\left( a + b \right)^2}=a+b}
   =
   x_1 + x_2 + y_1 + y_2
   \;\;\;\;  .
\end{equation}
\smallskip

\noindent If this sum is $O+I+\alpha+\beta$ or $x+x$, where $x\in\mathbb{F}_4$, then the Euclidean distance will be $O$. This means that the Euclidean distance on $\mathbb{F}_{16}$ does not satisfy the metric axiom $\mathbf{M2}$:\smallskip

\begin{equation} \label{FALL} 
   \exists (x_1,y_1), (x_2,y_2) \in \mathbb{F}_{16}\,
   \Big( 
      (x_1,y_1) \neq (x_2,y_2)
      \to
      d\left((x_1,y_1),(x_2,y_2)\right) = O 
   \Big)
   \;\;\;\;  .
\end{equation}
\smallskip

\noindent As an illustration, the Euclidean distance between any two distinct points on the line 11 or on the line 16 pictured in Fig. 1 is equivalent to zero. E.g., $d(P_{6},P_{11}) = I+\alpha+I+\alpha=O$ and $d(P_{10},P_{4})=I+\beta+\alpha+O=O$.\bigskip

\noindent Consequently, the Euclidean distance on the plane $\mathbb{F}_{16}$ is not metric. Provided the metric Euclidean distance is the natural way of measuring physical length of a line segment between two arbitrary points, it can be concluded that there is no notion of physical distance which can be defined everywhere on the plane $\mathbb{F}_4 \times \mathbb{F}_4$.\bigskip

\noindent Let $F(P)$ be a real function defined in each point $P=(x,y)$ on the plane $\mathbb{F}_4 \times \mathbb{F}_4$. The difference between two points is known as their delta, $\Delta P$, while the function difference, $\Delta F(P)$, divided by the point difference $\Delta P$ is known as “difference quotient'' \cite{Grossmann}:\smallskip

\begin{equation}  
   \frac{\Delta F(P)}{\Delta P}
   =
   \frac{F(P+\Delta P) - F(P)}{\Delta P}
   \;\;\;\;  .
\end{equation}
\smallskip

\noindent On condition that the difference between two distinct points $(x_1,y_1)$ and $(x_2,y_2)$ on $\mathbb{F}_4 \times \mathbb{F}_4$ is the Euclidean distance $d((x_1,y_1),(x_2,y_2))$, the difference quotient is given by the formula:\smallskip

\begin{equation}  
   \frac{\Delta F(P)}{\Delta P}
   =
   \frac{F(x_1,y_1) - F(x_2,y_2)}{x_1 + x_2 + y_1 + y_2}
   \;\;\;\;  .
\end{equation}
\smallskip

\noindent Based thereon, ${\Delta F(P)}/{\Delta P}$ is the slope of the secant line passing through the points with coordinates $\Big((x_1,y_1 ),F((x_1,y_1))\Big)$ and $\Big((x_2,y_2 ),F((x_2,y_2))\Big)$.\bigskip

\noindent Since for all $x$ in $\mathbb{F}_4$, $O \cdot x=O$, the division by $O$ must remain undefined. According to (\ref{FALL}), this implies that for some pairs of points on $\mathbb{F}_4 \times \mathbb{F}_4$, the slope ${\Delta F(P)}/{\Delta P}$ is undefined:\smallskip

\begin{equation}  
   F(x_1,y_1) \neq F(x_2,y_2)
   \to
   \frac{F(x_1,y_1) - F(x_2,y_2)}{O}
   \;\;\;\;  .
\end{equation}
\smallskip

\noindent Due to that, the difference quotient ${\Delta F(P)}/{\Delta P}$ cannot be considered as the mean value of the derivative of $F$ over the interval $[(x_1,y_1),(x_2,y_2)]$. One can infer then that the function $F$ does not have a derivative at the points of this interval because the function is not continuous there.\bigskip

\noindent In formal terms, the above implies that one cannot introduce a globally defined structure that makes possible differential calculus on the plane $\mathbb{F}_4 \times \mathbb{F}_4$. As a result, differential equations of motion cannot be applicable in the case of such a plane.\bigskip

\section{Emergent metricity of a plane geometry}  

\noindent The number $P(\mathcal{A})$ of points that constitute the region $\mathcal{A}$ can be presented as the number of elements in a finite set $S$, namely,\smallskip

\begin{equation}  
   P(\mathcal{A})
   =
   \mathrm{card}(S)
   \;\;\;\;  .
\end{equation}
\smallskip

\noindent In this way, using (\ref{SCOOR}) one gets\smallskip

\begin{equation}  
   \mathrm{card}\left( \mathcal{C}(\mathcal{A}) \right)
   =
   2^{P(\mathcal{A})}
   =
   2^{\,\mathrm{card}(S)}
   \;\;\;\;  .
\end{equation}
\smallskip

\noindent At the same time, since S is finite, there is a bijection from $S$ to the set of those natural numbers that are less than some specific natural number $n=\mathrm{card}(S)$, namely,\smallskip

\begin{equation}  
   f\!:\,\,
   S\to\{ 1,\dots, n \}
   \;\;\;\;  .
\end{equation}
\smallskip

\noindent For the region $\mathcal{A}$ of a macroscopic scale $\ell \in [1,10^{24}]$ m, or $\ell / \ell_P \in [10^{35},10^{59}]$, this number, i.e.,\smallskip

\begin{equation}  
   n
   =
   P(\mathcal{A})
   \sim
   \frac{\ell^2}{\ell_P^2}
   \;\;\;\;  ,
\end{equation}
\smallskip

\noindent is greater than $10^{70}$. In that instance, it can be believed that $n=\infty$ and so $S$ can be considered to be the set of all natural numbers $\mathbb{N}$. Symbolically,

\begin{equation}  
   \mathrm{card}\left( \mathcal{C}(\mathcal{A}) \right)
   \underset{\ell \,\gg\, \ell_P}{\longrightarrow}
   2^{\,\mathrm{card}(\mathbb{N})}
   \;\;\;\;  .
\end{equation}
\smallskip

\noindent By Cantor-Bernstein-Schroeder theorem \cite{Weisstein}, $2^{\,\mathrm{card}(\mathbb{N})} =\mathrm{card}(\mathbb{R})$. Thus, when $\ell \gg \ell_P$, the cardinality of the Cartesian coordinate system for $\mathcal{A}$ becomes the cardinality of the continuum:\smallskip

\begin{equation} \label{LIM} 
   \mathrm{card}\left( \mathcal{C}(\mathcal{A}) \right)
   \underset{\ell \,\gg\, \ell_P}{\longrightarrow}
   \mathrm{card}(\mathbb{R})
   \;\;\;\;  .
\end{equation}
\smallskip

\noindent Now, recall that in a field $\mathbb{F}_q$ of size $q=p^k$ (with a prime number $p$ and a positive integer $k$), adding $p$ copies of any element in $\mathbb{F}_q$ results in zero, specifically,\smallskip

\begin{equation}  
   \forall x \in \mathbb{F}_{p^k}\!:\,\,
   \underbrace{x+x+\dots+x}_{p\text{ copies}}=0
   \;\;\;\;  .
\end{equation}
\smallskip

\noindent Accordingly, one can say that the characteristic of the field $\mathbb{F}_q$ is $p$ \cite{Fraleigh}. If the above sum never reaches 0, the field $\mathbb{F}_q$ is said to have characteristic zero. For example, the field $\mathbb{R}$ consisting of all real numbers has characteristic 0.\bigskip

\noindent In keeping with (\ref{LIM}), a field having the characteristic 2 at $\ell \sim \ell_P$ becomes a field of characteristics 0 when $\ell \gg \ell_P$. Given that the metric axiom $\mathbf{M2}$ may hold true in fields of characteristics zero but breaks down in fields of characteristics 2, one can infer that a geometry of a plane seen as a metric at macroscopic scales $\ell \gg \ell_P$ ceases to be such in a scale $\ell \sim \ell_P$ whereat the Euclidean distance stops being metric. For that reason, the Planck length $\ell_P$ can be thought about as a length which is smaller than all possible physical distances.\bigskip

\section{Conclusion}  

\noindent At this point, the solution of the trans-Planckian problem along the lines of a finite geometry can be formulated in the following way.\bigskip

\noindent Seeing that there is no notion of physical distance defined across a finite field $\mathbb{F}_q$ of characteristics 2, the Einstein field equations -- the set of nonlinear partial differential equations -- cannot be given meaning at scales $\ell \sim \ell_P$, let alone $\ell < \ell_P$. The same applies to differential equations of any quantum field theory at $\ell \lesssim \ell_P$. Thus, a junction between general relativity and quantum mechanics is unable to render gravitational singularities: Neither of the theories has meaning in a scale $\ell \lesssim \ell_P$.\bigskip

\noindent This conclusion may serve as corroboration of the conjecture about radical modifications of the physical laws below the Planck length.\bigskip

\bibliographystyle{References}

\bibliography{Minimal_Ref}

\begin{thebibliography}{10}
\expandafter\ifx\csname urlstyle\endcsname\relax
  \providecommand{\doi}[1]{doi:\discretionary{}{}{}#1}\else
  \providecommand{\doi}{doi:\discretionary{}{}{}\begingroup
  \urlstyle{rm}\Url}\fi

\bibitem{Mead}
C.~Alden Mead.
\newblock {Observable Consequences of Fundamental-Length Hypotheses}.
\newblock \emph{Phys. Rev.}, 143:990--1005, 1966.

\bibitem{Adler}
Ronald~J. Adler and David~I. Santiago.
\newblock {On gravity and the uncertainty principle}.
\newblock \emph{Modern Physics Letters A}, 14(20):1371--1381, 1999.

\bibitem{Hossenfelder}
Sabine Hossenfelder.
\newblock {Minimal Length Scale Scenarios for Quantum Gravity}.
\newblock \emph{Living Reviews in Relativity}, 16(2):1--90, 2013.

\bibitem{Hooft}
Gerard ’t~Hooft.
\newblock {Black holes and the dimensionality of space-time}.
\newblock In Oskar Klein and Ulf Lindström, editors, \emph{{Proceedings of the
  Symposium “The Oskar Klein Centenary”, 19-21 Sept. 1994, Stockholm, Sweden}},
  pages 122--137. {World Scientific}, 1995.

\bibitem{Susskind}
Leonard Susskind.
\newblock {The World as a Hologram}.
\newblock \emph{J. Math. Phys.}, 36:6377--6396, 1995.

\bibitem{Bigatti}
Daniela Bigatti and Leonard Susskind.
\newblock {TASI lectures on the Holographic Principle}.
\newblock {https://arxiv.org/abs/hep-th/0002044}, Feb. 2000.

\bibitem{Salecker}
H.~Salecker and E.~P. Wigner.
\newblock {Quantum limitations of the measurement of space-time distances}.
\newblock \emph{Phys. Rev.}, 109:571--577, 1958.

\bibitem{Bousso}
Raphael Bousso.
\newblock {The holographic principle}.
\newblock \emph{Reviews of Modern Physics}, 74:825--874, 2002.

\bibitem{Jacobson}
Theodore Jacobson.
\newblock {Black-hole evaporation and ultrashort distances}.
\newblock \emph{Phys. Rev. D}, 44(6):1731--1739, 1991.

\bibitem{Brout}
R.~Brout, S.~Massar, R.~Parentanit, and Ph. Spindel.
\newblock {Hawking radiation without trans-Planckian frequencies}.
\newblock \emph{Phys. Rev. D}, 52(8):4559--4568, 1995.

\bibitem{Kiefer}
Claus Kiefer.
\newblock {Conceptual Problems in Quantum Gravity and Quantum Cosmology}.
\newblock \emph{ISRN Mathematical Physics}, 509316:1--18, 2013.

\bibitem{Bezdek}
K{\'{a}}roly Bezdek.
\newblock \emph{{Classical Topics in Discrete Geometry}}.
\newblock {Springer}, New York, NY, 2010.

\bibitem{Mullen13}
Gary~L. Mullen and Daniel Panario.
\newblock \emph{{Handbook of Finite Fields}}.
\newblock {Chapman and Hall/CRC}, New York, 2013.

\bibitem{Mullen07}
Gary~L. Mullen and Carl Mummert.
\newblock \emph{{Finite Fields and Applications}}.
\newblock {American Mathematical Society}, 2007.

\bibitem{Boros}
Endre Boros, Zoltan Füiredi, and L.~M. Kelly.
\newblock {On Representing Sylvester-Gallai Designs}.
\newblock \emph{Discrete and Computational Geometry}, 4:345--348, 1989.

\bibitem{Burago}
Dmitri Burago, Yuri Burago, and Sergei Ivanov.
\newblock \emph{{A Course in Metric Geometry}}.
\newblock {American Mathematical Society}, 2001.

\bibitem{Grossmann}
Christian Grossmann, Hans-Görg Roos, and Martin Stynes.
\newblock \emph{{Numerical Treatment of Partial Differential Equations}}.
\newblock {Springer-Verlag}, Berlin Heidelberg, 2007.

\bibitem{Weisstein}
Eric~W. Weisstein.
\newblock {Schröder-Bernstein Theorem. From MathWorld--A Wolfram Web Resource}.
\newblock {https://mathworld.wolfram.com/Schroeder-BernsteinTheorem.html},
  2023.

\bibitem{Fraleigh}
John~B Fraleigh and Neal Brand.
\newblock \emph{{First Course in Abstract Algebra, A}}.
\newblock {Pearson}, 2021.

\end{thebibliography}

\end{document}